\documentclass{iaus}
\usepackage{graphicx}

\title[AGN properties of NGC 7582] 
{The AGN properties of the \\ starburst galaxy NGC 7582.}

\author[T.V. Ricci et al.]   
{T.V. Ricci$^{1\dagger}$, J.E. Steiner$^1$, R.B. Menezes$^1$, A. Garcia-Rissmann$^2$ \and R. Cid Fernandes$^3$}

\affiliation{$^1$Instituto de Astronomia, Geof\'isica e Ci\^encias Atmosf\'ericas, Universidade de S\~ao Paulo, \\
Rua do Mat\~ao, 1226, S\~ao Paulo - SP, Brasil \\[\affilskip]
$^2$Gemini Observatory \\ [\affilskip] $^3$Universidade Federal de Santa Catarina \\ $^{\dagger}${\tt tiago@astro.iag.usp.br}}

\pubyear{2009}
\volume{267}  
\pagerange{119--126}
\jname{Co-Evolution of Central Black Holes and Galaxies}
\editors{B.M.\ Peterson, R.S.\ Somerville, \& T.\ Storchi-Bergmann, eds.}
\begin{document}

\maketitle

\begin{figure}[h]
\begin{center}
\begin{minipage}[b]{0.45\linewidth}
\includegraphics[width=1.2 \columnwidth,angle=0]{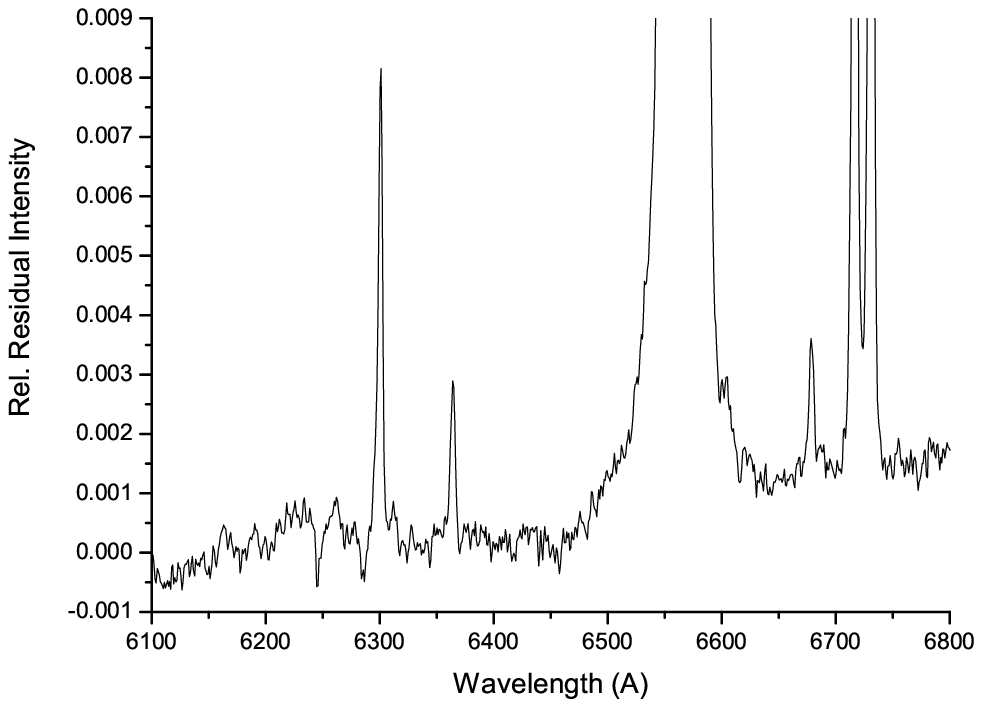}
\caption{Spectra extracted from the region of the AGN.}
\label{fig1}
\end{minipage} \hfill
\begin{minipage}[b]{0.45\linewidth}
\includegraphics[width=1.2 \columnwidth,angle=0]{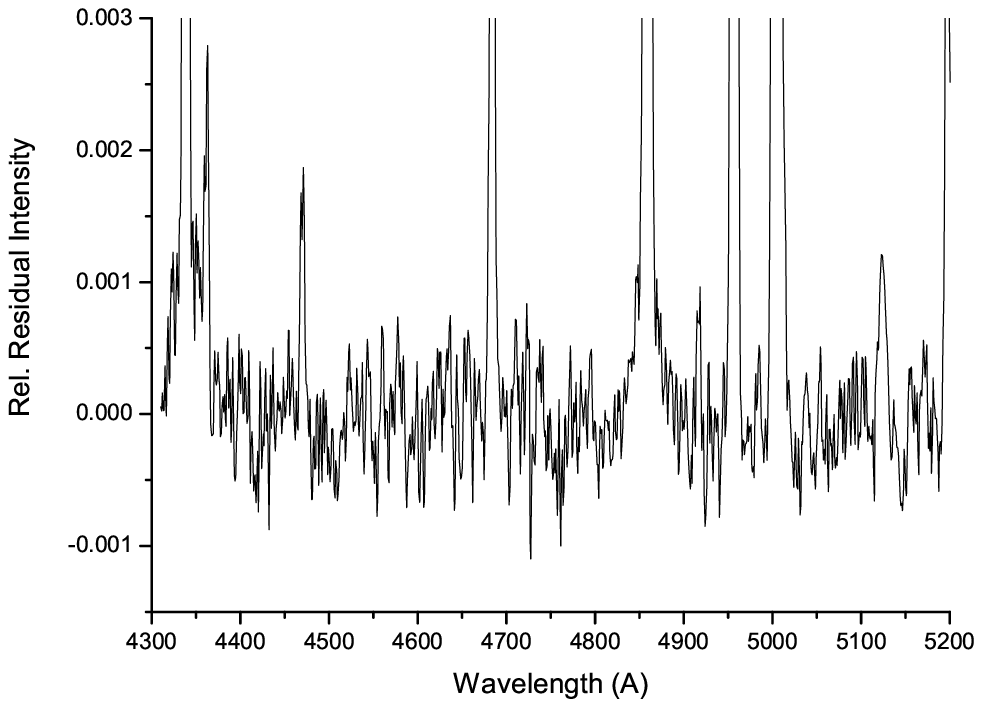}
\caption{Spectra extracted from the region of the ionization cone.}
\label{fig2}
\end{minipage}
\end{center}
\end{figure}

NGC 7582 was identified as a Starburst galaxy in the optical \cite[(Veron et al. 1981)]{Veron et al.(1981)} but its X-Ray emission is typical of a Seyfert 1 galaxy \cite[(Ward et al. 1978)]{Ward et al.(1978)}. We analyzed a datacube of this object obtained with the GMOS-IFU on the Gemini-South telescope. After a subtraction of the stellar component using the {\sc starlight}  code \cite[(Cid Fernandes et al. 2005)]{Cid Fernandes et al. (2005)}, we looked for optical signatures of the AGN. We detected a broad $H\alpha$ component (figure \ref{fig1}) in the source where \cite[Bianchi et al.(2007)]{Bianchi et al.(2007)} identified the AGN in an HST optical image. We also found a broad $H\beta$ feature (figure \ref{fig2}), but its emission reveals a extended source. We suggest that it is the light of the AGN scattered in the ionization cone. We propose that NGC 7582 is a Seyfert 1 galaxy. A number of other ``hot-spots'' and Wolf-Rayet features were also identified.

\end{document}